\documentstyle[aaspp4]{article}
\input psfig.sty

\def\lea{\mathrel{<\kern-1.0em\lower0.9ex\hbox{$\sim$}}}
\def\gea{\mathrel{>\kern-1.0em\lower0.9ex\hbox{$\sim$}}}

\def\ltap {\mathrel{\hbox{\rlap{\lower.55ex \hbox {$\sim$}} \kern-.3em
                   \raise.4ex \hbox{$<$}}}}
\def\gtap {\mathrel{\hbox{\rlap{\lower.55ex \hbox {$\sim$}} \kern-.3em
                   \raise.4ex \hbox{$>$}}}}
\def\arcm               {$^{\prime}$} 
\def\arcs               {$^{\prime\prime}$} 

\def\bily               {${ {10^9}}$yr}
\def\cent               {$\omega$\thinspace Centauri}
\def\cen                {$\omega$\thinspace Cen}
\def\chis               {$\chi^2$}
\def\conc               {$c$ = log ($r_t/r_c$)}
\def\deg                {$^\circ$}
\def\etal               {et\thinspace al.}
\def\kilm               {${ {10^3 \mathcal{M}_{\odot}}}$}

\def\kms                {km\thinspace s$^{-1}$}
\def\gone               {Mayall~II $\equiv$~G1}
\def\Mass               {$\mathcal{M}$}
\def\Mhr                {${\mathcal{M}}_{tot}^{hr}$}

\def\mlv                {${\mathcal M}/L_V$}

\def\milm               {${ {10^6 \mathcal{M}_{\odot}}}$}
\def\mily               {${ {10^6}}$yr}
\def\Mtot               {${\mathcal{M}}_{tot}$}

\def\muov               {$\mu$(0,$V$)}

\def\Mv                 {$M_V$}

\def\Mvtot              {$M_V^{tot}$}
\def\msun               {$M_{\odot}$}

\def\pmm                {$\pm$}
\def\rhoc               {$\rho_{c}$}
\def\rhoo               {$\rho_{\circ}$}
\def\rhoh               {$\rho_{h}$}
\def\rhot               {$\rho_{t}$}
\def\ra                 {$r_a$}
\def\rc                 {$r_c$}
\def\rh                 {$r_h$}
\def\rt                 {$r_t$}
\def\sobs               {$\sigma_{obs}$}

\def\sigo               {$\sigma$(0)}

\def\sigpo              {$\sigma_p$(0)}

\def\trh                {t$_{rh}$}
\def\tro                {t$_{r,\circ}$}
\def\tuca               {47\thinspace Tucanae}

\def\vs                 {$v_{s}$}
\def\wo                 {$W_{\circ}$}
\def\x                  {$\times$}

\lefthead{Meylan \etal}  \righthead{  Mayall~II $\equiv$~G1:  Globular
Cluster or Dwarf Galaxy~? }

\begin{document}

\title{Mayall~II $\equiv$~G1 in M31: \\
Giant Globular Cluster or Core of a Dwarf Elliptical Galaxy?~$^{1,2}$ }

\author{ G. Meylan$^{3}$ }
\affil{  Space Telescope Science Institute, 3700 San Martin Drive, \\
         Baltimore, MD 21218, U.S.A.,  email: gmeylan@stsci.edu, \\
                                 and  \\
         European Southern Observatory, Karl-Schwarzschild-Str. 2, \\
         D-85748 Garching bei M\"unchen, Germany. }

\author{ A. Sarajedini }
\affil{  University of Florida, Astronomy Department, 
         Gainesville, FL 32611-2055, U.S.A. \\
         email: ata@astro.ufl.edu }

\author{ P. Jablonka }
\affil{  DAEC-URA 8631, Observatoire de Paris-Meudon, \\
         Place Jules Janssen, F-92195 Meudon, France.~ 
         email: jablonka@daec.obspm.fr }

\author{ S. G. Djorgovski }
\affil{  Palomar Observatory, MS 105--24, Caltech, 
         Pasadena, CA 91125, U.S.A. \\
         email: george@oracle.caltech.edu }

\author{ T. Bridges }
\affil{  Anglo-Australian Observatory, PO Box 296, 
         Epping, NSW 2121, Australia. \\
         email: tjb@aaoepp.aao.gov.au }

\author{ R. M. Rich }
\affil{  UCLA, Physics \& Astronomy Department, Math-Sciences 8979, \\
         Los Angeles, CA 90095-1562, U.S.A.~
         email: rmr@astro.ucla.edu }

 $^1$  Based in  part on  observations made  with the  NASA/ESA Hubble
         Space  Telescope,  obtained at  the  Space Telescope  Science
         Institute,   which  is   operated  by   the   Association  of
         Universities  for  Research in  Astronomy,  Inc., under  NASA
         contract NAS 5-26555.   These observations are associate with
         proposal IDs 5907 and 5464.

 $^2$  Based  in  part  on  observations  obtained  at  the  W.M.~Keck
         Observatory,  which  is operated  jointly  by the  California
         Institute of Technology and the University of California.

 $^3$ Affiliated with the  Astrophysics Division of the European Space
         Agency, ESTEC, Noordwijk, The Netherlands.


\begin{abstract}

Mayall~II  $\equiv$~G1  is  one  of the  brightest  globular  clusters
belonging  to M31, the  Andromeda galaxy.   Our observations  with the
Wide  Field  and  Planetary  camera  WFPC2 onboard  the  Hubble  Space
Telescope  (HST) provide  photometric data  for the  $I$~vs.~$V-I$ and
$V$~vs.~$V-I$  color-magnitude   diagrams.   They  reach   stars  with
magnitudes  fainter than  $V$  = 27  mag,  with a  well populated  red
horizontal branch at about $V$ = 25.3 mag.

From model  fitting, we  determine a rather  high mean  metallicity of
[Fe/H] =  --0.95 \pmm\ 0.09, somewhat  similar to \tuca.   In order to
determine our true measurement  errors, we have carried out artificial
star  experiments.  We  find  a larger  spread  in $V-I$  than can  be
explained  by the  measurement errors,  and  we attribute  this to  an
intrinsic metallicity dispersion amongst the  stars of G1; this may be
the consequence of  self-enrichment during the early stellar/dynamical
evolutionary phases of this cluster.
%
%
So  far, only  \cent, the  giant Galactic  globular cluster,  has been
known  to   exhibit  such  an  intrinsic   metallicity  dispersion,  a
phenomenon certainly related to the  the deep potential wells of these
two star clusters.

We determine, from the  same HST/WFPC2 data, the structural parameters
of G1.  Its surface brightness profile provides its core radius \rc\ =
0.14\arcs\  =  0.52~pc, its  tidal  radius  \rt\  $\simeq$ 54\arcs\  =
200~pc,  and  its concentration  \conc\  $\simeq$  2.5.   Such a  high
concentration  indicates the  probable  collapse of  the  core of  G1.
KECK/HIRES observations provide the central velocity dispersion \sobs\
= 25.1~\kms, with \sigpo\ = 27.8~\kms\ once aperture corrected.

Three  estimates of the  total mass  of this  globular cluster  can be
obtained.  The King-model mass is  \Mass$_K$ = 15 $\times$ \milm\ with
\mlv\  $\simeq$  7.5, and  the  Virial  mass  is \Mass$_{Vir}$  =  7.3
$\times$ \milm\ with \mlv\ $\simeq$ 3.6.  By using a King-Michie model
fitted  simultaneously  to  the  surface brightness  profile  and  the
central   velocity  dispersion  value,   mass  estimates   range  from
\Mass$_{KM}$ = 14 $\times$ \milm\ to 17 $\times$ \milm.

Although  uncertain, all  of these  mass estimates  make G1  more than
twice as massive as \cent, the most massive Galactic globular cluster,
whose mass is also uncertain by about a factor of 2.  G1 is not unique
in M31: at least 3 other  bright globular clusters of this galaxy have
velocity  dispersions \sobs\  larger than  20~\kms,  implying probably
similar large masses.

Such large  masses relate  to the metallicity  spread whose  origin is
still unknown (either  self-enrichment, an inhomogeneous proto-cluster
cloud, or remaining  core of a dwarf galaxy).  Let  us consider for G1
(see   Table~1)  the  four   following  parameters:   central  surface
brightness  \muov\  =  13.47~mag~arcsec$^{-2}$,  core  radius  \rc\  =
0.52~pc, integrated absolute visual  magnitude \Mv\ = --10.94 mag, and
central velocity  dispersion \sigo\  = 28~\kms.  When  considering the
positions of G1  in the different diagrams defined  by Kormendy (1985)
using the  above four  parameters, G1 always  appears on  the sequence
defined  by globular  clusters,  and definitely  away  from the  other
sequences defined by elliptical galaxies, bulges, and dwarf spheroidal
galaxies.  The same is true for \cent.

Little is known about the  positions, in these diagrams, of the nuclei
of nucleated dwarf elliptical galaxies, which could be the progenitors
of the most massive globular  clusters.  The above four parameters are
known only for the nucleus of one dwarf elliptical, viz., NGC~205, and
put this object, in the Kormendy's  diagram, close to G1, right on the
sequence of globular clusters.  This does not prove that all (massive)
globular clusters  are the remnant cores of  nucleated dwarf galaxies.

At  the moment,  only  the anti-correlation  of  metallicity with  age
recently observed in \cent\ suggests that this cluster enriched itself
over a  timescale of about  3~Gyr.  This contradicts the  general idea
that all the stars in a globular cluster are coeval, and may favor the
origin of \cent\ as being the remaining core of a larger entity, e.g.,
of a former nucleated dwarf  elliptical galaxy.  In any case, the very
massive globular  clusters, by the  mere fact that their  large masses
imply complicated stellar and dynamical evolution, may blur the former
clear (or  simplistic) difference between globular  clusters and dwarf
galaxies.

%
%
%

\end{abstract}

\keywords{
galaxies: dwarf, evolution, formation, Local Group, star clusters;  
Galaxy: globular clusters: general 
Galaxy: globular clusters: individual (\cent, Mayall~II $\equiv$~G1}


\section{Introduction}

From  both stellar  population and  stellar dynamics  points  of view,
globular  clusters  represent a  very  interesting  family of  stellar
systems.   They are ancient  building blocks  of galaxies  and contain
unique information about old  stellar populations and their formation.
Some fundamental dynamical processes take place in their cores on time
scales  shorter than  the  age  of the  universe,  offering us  unique
laboratories for learning  about two-body relaxation, mass segregation
from equipartition of energy, stellar collisions, stellar mergers, and
core collapse.  See Meylan \& Heggie (1997) for a general review.

In  our Galaxy,  globular clusters  span  a wide  range of  properties
(Djorgovski \&  Meylan 1994).  For example,  their integrated absolute
magnitudes and total masses range from \Mvtot\ = --10.1 and \Mtot\ = 5
\x\ \milm\ (Meylan \etal\ 1994,  1995) for the giant Galactic globular
cluster \cent\ down to \Mvtot\  = --1.7 and \Mtot\ $\simeq$ \kilm\ for
the Lilliputian Galactic globular cluster AM-4 (Inman \& Carney 1987).
AM-4,  located at  $\simeq$ 26~kpc  from  the Galactic  centre and  at
$\simeq$  17~kpc above  the Galactic  plane,  does not  belong to  the
Galactic disk, and consequently cannot be considered to be an old open
cluster.   The  uncertainties  on  the  above  total  mass  estimates,
typically  as large  as 100\%,  do  not alter  the fact  that, in  our
Galaxy, the  individual masses of  globular clusters range  over three
orders  of magnitude.   It  is not  known  to what  extent these  mass
differences  are  ``congenital''  or  due  to  subsequent  pruning  by
dynamical evolution.

With its approximately 450 members (Barmby et al.  2000), the globular
cluster system of  M31, the Andromeda galaxy, is  about three times as
rich  as the  Galactic one,  and is  among the  most  studied globular
cluster  systems in  external  galaxies (Harris  1991).  However,  our
knowledge  comes   mainly  from  the   integrated  photometric  and/or
spectroscopic properties  of these  clusters (e.g., Reed  \etal\ 1994,
Barmby  et al.   2000).  It  is essentially  since the  advent  of the
Hubble  Space  Telescope  (HST),  with its  post-refurbishment  camera
WFPC2,  that  we  resolve  clearly  some  of  the  M31  clusters  into
individual  stars,  giving  access  to their  general  morphology  and
structural parameters (e.g., Fusi  Pecci \etal\ 1994, Grillmair \etal\
1996), and  providing their Color-Magnitude  Diagrams (CMDs), reaching
magnitudes  fainter than  the  Horizontal Branch  (e.g., Ajhar  \etal\
1996, Fusi Pecci \etal\ 1996, Rich \etal\ 1996, Jablonka \etal\ 2000).
Correlations between structural,  photometric and dynamical parameters
have  been investigated for  21 globular  clusters in  M31 (Djorgovski
\etal\ 1997).

The brightest globulars in M31 are brighter than our Galactic champion
\cent.  Among  these giants is  \gone\ (Rich \etal\ 1996),  a globular
cluster  so bright that,  like \cent,  it has  been considered  as the
possible  remaining core  of a  former dwarf  elliptical  galaxy which
would have lost most of it envelope through tidal interaction with its
host galaxy  (Meylan  \etal\ 1997, 2000, Meylan 2000).  We  present in
this paper a detailed photometric and dynamical study of G1.

This  paper is  structured as  follows: Section~2  gives  some general
information  about  Mayall~II  $\equiv$~G1,  Section~3  describes  the
observations, Section~4 gives  the CMD of G1 and  discusses the spread
in  metallicity observed  among the  stars  of the  red giant  branch,
Section~5  presents  its  ellipticity,  position  angle,  and  surface
brightness profile, Sections~6 and~7 give the results from the various
mass  estimators.   All  results   are  summarized  and  discussed  in
Section~8.


\section{Mayall~II $\equiv$~G1, a luminous globular cluster in M31}

The  globular cluster G1  belongs to  our companion  galaxy, Andromeda
$\equiv$ M31.   Resolved with  difficulty from the  ground (Djorgovski
1988, Heasley \etal\ 1988, Bendinelli \etal\ 1990), this huge swarm of
stars appears  as a bright  flattened star cluster when  observed with
the Wide  Field and  Planetary camera WFPC2  onboard the  Hubble Space
Telescope (HST) (see Fig.~1).  The integrated visual magnitude of this
cluster, $V$ = 13.48 mag,  corresponds to an absolute visual magnitude
M$_V$  = --10.94  mag, with  $E(B-V)$ =  0.06 and  a  distance modulus
$(m-M)_{M31}$ = 24.43 mag, implying  a total luminosity of about $L_V$
$\sim$  2  $\times$ $10^6  L_{\odot}$  (Rich  \etal\ 1996,  Djorgovski
\etal\ 1997).

The coordinates of G1,  
when compared to the coordinates of the center of M31 (see Table~1), 
place it at a projected distance of about 3\deg, i.e. 40~kpc, from the
center of M31.  This rather  large projected distance is comparable to
the  distance  between our  Galaxy  and  the  Large Magellanic  Cloud.
Nevertheless, both  color-magnitude diagrams and  radial velocities of
G1 ($V_r {\rm  (G1)} = - 331 \pm\ 24~  {\rm km\thinspace s}^{-1}$) and
M31 ($V_r {\rm (M31)} = -  300 \pm\ 4~ {\rm km\thinspace s}^{-1}$ from
21cm HI  line and $V_r {\rm (M31)}  = - 295 \pm\  7~ {\rm km\thinspace
s}^{-1}$ from  optical lines), completely  support the idea  that this
cluster belongs to the globular  cluster system of M31.  The values of
the most  important general parameters describing G1  are displayed in
Table~1.


\section{Observations}

We use our observations obtained with the Planetary Camera (PC) of the
HST/WFPC2, in the framework of  a programme (PI Pascale Jablonka, ID =
5907) aiming at studying star  clusters and stellar populations in M31
(see Jablonka \etal\ 1999, Jablonka  \etal\ 2000).  The PC pixel scale
is 0.045\arcs pix$^{-1}$.   The four images of G1,  taken with each of
the F555W ($V$) and F814W  ($I$) filters, have total integration times
equal   to  500~+~500~+~600~+~600~=~2,200  seconds   in  $V$   and  to
400~+~400~+~500~+~500~=~1,800  seconds in  $I$.   Because of  possible
non-linearity in the  bright concentrated core of G1,  our rather deep
exposures are  supplemented with  some shorter exposures  from another
programme (PI R. Michael Rich, ID = 5464).  See Rich \etal\ (1996).

Fig.~1  displays  an area  of  31.5\arcs~$\times$~31.5\arcs\ from  the
original PC frames centered on the cluster.  This image is a composite
of all our $V$ and $I$ frames and provides a genuine indication of the
relative  colors  of the  stars.   Although  completely resolved,  the
cluster  appears   extremely  compact,  with  a   very  steep  surface
brightness profile and an extremely bright and crowded core.

We determine the photometry using  one of the presently best available
algorithms for performing stellar  photometry in crowded fields, viz.,
the ALLFRAME  procedure developed by P.  Stetson  (1994).  ALLFRAME is
run   on   700~$\times$~700-pixel   (31.5\arcsec~$\times$~31.5\arcsec)
sub-areas  of the  original (36\arcsec~$\times$~36\arcsec)  PC frames,
avoiding the egdes of the frame and masking the two areas disturbed by
the two bright  foreground stars (Fig.~1).  As ALLFRAME  is now widely
known and since our use of  it is already described in Jablonka \etal\
(1999), here  we mainly  focus on the  results.  Given the  very large
projected  distance between G1  and the  core of  M31 (40~kpc),  it is
worth mentioning that the number of M31 field stars in our PC field is
negligible when compared to the number of G1 stars.


\section{The Color-Magnitude Diagram of Mayall~II $\equiv$~G1}

Fig.~2 displays the two color-magnitude diagrams (CMDs) of G1, each of
them  containing  the same  4903  stars.   The  left panel  shows  the
$V$~vs.~$V-I$ CMD,  while the  right panel displays  the $I$~vs.~$V-I$
CMD.   The  brightest stars  at  $V$$\sim$22.5  have  color errors  of
$\sim$0.03, and  stars at  the level  of the HB  have color  errors of
$\sim$0.15 mag.

A first  CMD of G1, reaching  stars below the  horizontal branch (HB),
was published  by Rich \etal\ (1996)  based on HST  Cycle~4 data, with
total exposure  times of 1,600 seconds  in F555W and  1,200 seconds in
F814W.   Our exposure  times are  about 40\%  larger in  $V$  and 50\%
larger in $I$.  It is why, with the use of different methods/softwares
in  the photometric  analysis, we  can reach  stars 0.5~mag  and 1~mag
fainter in $V$ and $I$, respectively.

%
%
%

These two CMDs show a  relatively shallow RGB  and a horizontal branch
(HB)  populated predominantly redward   of  the  RR~Lyrae  instability
strip.  Both of these features suggest that G1  is a rather metal-rich
stellar system. However, the CMD  also reveals a blueward extension to
the  red HB clump  composed of a small  number of stars.  All three of
these populations were also noted by Rich \etal\   (1996) in their CMD
of G1.  In  particular, Rich \etal\  (1996)  pointed out that the blue
extension to the    HB could possibly  be   the result of a   chemical
abundance spread  in G1.  Indeed, the  RGB  does display a potentially
significant  color width.  The  statistical significance of this width
is addressed below.

We note  that the morphology  of G1 HB  is more reminiscent of  the HB
morphology  observed in  the dwarf  spheroidal galaxy  Andromeda~I (Da
Costa  \etal\ 1996)  than  in the  globular  cluster \tuca\  (Vazdekis
\etal\ 2001).  This  fits some of the conclusions  of this work, which
unveiled some  characteristics of the  stellar population of  G1, more
typical  of dwarf  galaxies than  of  our classical  view of  globular
clusters.
 
In order to measure the magnitude of the HB, we construct a luminosity
function of the data  and fit a Gaussian curve  to the  most prominent
peak  in this luminosity  function.   This procedure  yields  $V(HB) =
25.34  ~\pm~  0.07$.  The quoted   error is the   result of adding, in
quadrature,  estimated  errors of \pmm~0.05   in  the determination of
$V(HB)$ and   \pmm~0.05 in  the photometric   zeropoint.   This  is in
excellent agreement with  the work of  Rich \etal\ (1996) who obtained
$V(HB) = 25.32 ~\pm~  0.05$.  To estimate  the metallicity of G1  in a
way that is   independent   of  the  photometric zeropoint  and    the
reddening, we rely   upon the  slope of the    RGB as calibrated    by
Sarajedini \etal\ (2000).   Utilizing their measurement  technique and
calibration leads to  a value of [Fe/H]  = $-0.95 ~\pm~ 0.09$ for  the
mean metal abundance of G1 on the  scale of Zinn  \& West (1984). This
abundance lies between the results of Rich  \etal\ (1996) who obtained
[Fe/H] $\sim$ --0.7 on the scale of Zinn \& West (1984), a value close
to that of  \tuca, and  those of  Bonoli  \etal\ (1987) and Brodie  \&
Huchra  (1990)  who obtained [Fe/H]  $\sim$  --1.2.   In addition, our
abundance is in  accord with the estimate  of Stephens  et al.  (2001)
based  on the $V-K(RGB)$  of G1; they  find [Fe/H] = $-0.9 ~\pm~ 0.2$.
Lastly,  we can utilize the mean  RGB color  along with the calculated
metal  abundance and   Equ.~1 of   Sarajedini (1994)   to compute  the
reddening of G1; we find $E(V-I) = 0.05 ~\pm~ 0.02$.

Let us  return now to the  RGB color width apparent in  the CMD  of G1
(see Fig.~2).    If   this feature is  significant,  i.e.   not caused
entirely  by the photometric  errors, then we  can argue strongly that
there is a metallicity dispersion in G1.  One method used to test this
is  to conduct artificial  star experiments  in  order to estimate the
true measurement error.  To begin  with, we select stars located along
an RGB fiducial sequence.  For each of two trials, we
select 210  stars along this fiducial  sequence and  place them on the
original PC1 frames under the  constraint that no two artificial stars
be within two PSF radii of each  other.  The resultant images are then
reduced with the same procedure as the original PC images.

The  filled circles in  Fig.~3 represent  the original  magnitudes and
colors of  the 420 artificial stars  while the open  circles are their
recovered  values.   We are  interested  in  the  color width  of  the
artificial stars  and how  this compares with  the observed  RGB color
width.  The open  circles in the lower panel of  Fig.~3 show the color
histogram of stars  located 1.8 \pmm\ 0.25 magnitudes  above the HB of
G1.  This  location is  chosen because it  minimizes the  influence of
asymptotic giant branch stars (Da Costa \& Armandroff 1990; Geisler \&
Sarajedini 1999).  The filled circles represent the color histogram of
the artificial  stars around the fiducial sequence.   Gaussian fits to
these distributions  yield $\sigma_{obs} = 0.144 ~\pm~  0.010$ for the
observed width  and $\sigma_{err} =  0.037 ~\pm~ 0.004$ for  the width
due to photometric errors.  Subtracting these quantities in quadrature
gives $\sigma_{int}  = 0.139 ~\pm~  0.011$ for the intrinsic  width of
the RGB. We note in  passing that these artificial star experiments do
not provide a complete and total assessment of the photometric errors.
Other   sources  of   error,   e.g.,  such   as  residual   flat-field
non-uniformities and  residual dark current,  are not included  in our
photometric error estimates.

%
%

If we assume then that the intrinsic photometric width we calculate is
due  entirely  to  a  metallicity   dispersion  in  G1,  what  is  the
corresponding range in [Fe/H]~?  To estimate this quantity, we turn to
the standard RGB sequences of  Da Costa \& Armandroff (1990), which we
use to construct  a relation between [Fe/H] and  $(V-I)_0$ at 1.8 mags
above  the  HB  of the  six  standard  clusters  in that  study.   The
resultant relation is quadratic, which  means that the [Fe/H] range it
implies for G1 depends on the reddening we assume.  The relation is:
$$ 
{\rm [Fe/H] } = -18.648 + 23.846(V-I)_0 - 7.906(V-I)_0^2 \eqno(1)
$$
with $(V-I)_0$  = 1.39 at $V(HB)  - 1.8$.  For example,  if $E(V-I)$ =
0.10, then  we infer a  1-$\sigma$ [Fe/H] dispersion  of $\sigma_{{\rm
[Fe/H]}}  =  ~\pm~  0.50$~dex;   whereas  if  $E(V-I)$  =  0.05,  then
$\sigma_{{\rm  [Fe/H]}}   =  ~\pm~   0.39$~dex.   In  any   case,  the
metallicity dispersion in G1 is genuine and significant.  In contrast,
we  applied the  above technique  to HST/WFPC2  photometry of  the M31
globular cluster G219 (Neill 2001).   These images were reduced in the
same manner  as those of G1  presented herein. We find  that the color
width of  its RGB is fully  consistent with the  photometric errors as
expected for a system with a negligible metallicity dispersion.

The fact that in their  HST/NICMOS study of G1, Stephens \etal\ (2001)
do not observe  any spread in metallicity does  neither contradict nor
infirm  our present  result. This  for  two reasons.   First, for  any
spread in [Fe/H], the corresponding  spread in color is twice as small
in the  infrared $(J-K)$ than  in the visible $(V-I)$.   Second, their
very small sample (they have about 200 stars while we have about 5,000
stars) would  certainly affect statistically their  measurement of any
spread in metallicity.

The only other  globular cluster known to exhibit  a significant metal
abundance  range is  $\omega$  Centauri, the  giant Galactic  globular
cluster   (see   Jurcsik  1998   for   a   compilation  of   abundance
measurements). Its  range in [Fe/H] is  roughly one dex  (Norris \& Da
Costa  1995), which  is quite  similar to  the range  inferred  by our
estimates for G1.


\section{ Ellipticity, Position Angle, and Surface Brightness Profile }

Surface photometry of the cluster is obtained from all available WFPC2
images, using the techniques  described by Djorgovski (1988).  We then
combine  the  profiles  extracted  from different  images,  using  the
shorter, unsaturated  exposures for the  central part of  the profile,
and  longer, higher  S/N  exposures  (in which  the  cluster core  was
saturated) for the outer parts of the profiles.

Table~2 gives the ellipticity $\epsilon$  = $1 - b/a$ and the position
angle $PA$ as  a function of the semi-major axis  $a$, using the stack
of all our  F555W ($V$) frames along with  short exposures obtained in
the same filter  and available in the STScI/HST  archives.  These data
are displayed  in Fig.~4.   The ellipticity varies  significantly with
the semi-major axis $a$, from  values smaller than $\epsilon$ = 0.1 in
the  innermost and  outermost parts  of the  cluster, but  with values
larger than $\epsilon$ = 0.2  between 0.7\arcs\ and 7\arcs, reaching a
maximum of $\epsilon$ = 0.3 at $a$ = 2.1\arcs. The mean ellipticity of
G1  is  $\epsilon$ $\simeq$  0.2.   The  position  angle $PA$  is  not
significantly  variable for  semi-major  axis values  $a$ larger  than
0.2\arcs.  There is no significant evidence for twist of isophotes.

Table~3  gives  the  surface  brightness $\mu_V$  and  integrated  $V$
magnitude as a function of the radius.  The 72 points of this observed
surface  brightness profile  are  displayed in  Fig.~5.   There is  no
observational evidence  of the presence of unbound  stars and/or tidal
tails surrounding  G1, simply by the  mere fact that we  would need to
reach  stars a  few  magnitudes fainter  than  the turnoff  to have  a
statistically significant sample of such escaping stars.

In  order  to  convert  the  observed  count rates  to  the  $V$  band
magnitudes,  we used  the  standard transformations  from Holtzman  et
al. (1995), assuming for the color  of the cluster $(B-V) = 0.84$ mag.
Integration of the combined profile  gives the total magnitude for the
cluster,  $V_{int} =  13.48 \pm  0.05$  mag, where  the net  estimated
zero-point uncertainty includes both  the errors due to the background
subtraction, and  the color transformation  (they are of  a comparable
magnitude).  This  is in an  excellent agreement with the  values from
van  den Bergh  (1969)  and Reed  et  al. (1994),  after applying  the
appropriate aperture and color corrections, which give 13.54 and 13.58
mag, respectively.   We note that these  ground-based measurements are
likely to  underestimate slightly the  cluster brightness, due  to the
removal of  the light  covered by the  bright foreground  stars, which
accounts for some  of the systematic difference here.   (None of these
numbers include the Galactic extinction corrections.)


\section{ Mass Estimators }

We have   at  our disposal  two  essential  observational  constraints
allowing the mass determination  of G1.  First, our new  determination
of  its  surface brightness  profile (see   Table~3  and  Fig.~5) from
HST/WFPC2 images, providing  essential structural parameters: the core
radius \rc\  =  0.14\arcs\ =  0.52~pc,   the half-mass radius \rh\   =
3.7\arcs\ =  14~pc, the tidal radius \rt\  $\simeq$ 54\arcs\ = 200~pc,
implying  a concentration  \conc\  $\simeq$ 2.5.  Second,  our already
published  central   velocity  dispersion   from   KECK/HIRES spectra,
providing  an observed velocity dispersion  \sobs\ = 25.1~\kms, and an
aperture-corrected  central velocity  dispersion \sigpo\ =  27.8~\kms\
(Djorgovski \etal\ 1997).

\subsection{ King model and Virial mass estimates }

Masses of dynamical systems  are difficult to evaluate, with different
methods providing  rather different  results, and the  scatter between
the  different  values is  generally  much  larger  than their  formal
uncertainties.  Consequently, it is worth presenting results from more
than  one  method,  thus  giving  a realistic  estimate  of  the  true
uncertainty.  We can  first  obtain simple  mass  estimates from  King
models and from the Virial theorem (see, e.g., Illingworth 1976).

The first  estimate, the King mass  \Mass$_K$, is given  by the simple
equation:
$$
{\mathcal{M}}_K ~=~ \rho_c r_c^3 \mu ~=~ 167~r_c \mu \sigma_{\circ}^2 \eqno(2)
$$
where the core radius \rc\ = 0.52~pc, the dimensionless quantity $\mu$
=  220 for \conc\    = 2.5  (King   1966),  and the  central  velocity
dispersion \sigpo\ = 27.8~\kms.   These values determine a  total mass
for the cluster of \Mtot\ = 15 $\times$  \milm\ with the corresponding
\mlv\ $\simeq$ 7.5 in solar units.

The second estimate, the  Virial mass \Mass$_{Vir}$,  is given by  the
simple equation:
$$
{\mathcal{M}}_{Vir} ~=~ 670~r_h  \sigma_{\circ}^2 \eqno(3)
$$
where  the  half-mass  radius   \rh\  =  14~pc  and  central  velocity
dispersion \sigpo\  = 27.8~\kms.  These values determine  a total mass
for the cluster of \Mtot\ = 7.3 $\times$ \milm\ with the corresponding
\mlv\ $\simeq$ 3.6 in solar units. The internal error of each of these
mass estimates amounts to about 10\%.

The large difference between these  two estimates is not particular to
the present cluster, but due  to the idiosyncrasies of each method and
typical of these two mass  estimators applied to any globular cluster.
See, e.g.,  Table~6 showing the  results of similar mass  estimates in
the case  of \cent, the  brightest and most massive  Galactic globular
cluster.

\section{ King-Michie model mass estimates }

The  two  above observational  constraints,  viz., surface  brightness
profile and central velocity dispersion, allow the use of a multi-mass
King-Michie model  as defined by  Gunn \& Griffin (1979).   See Meylan
(1987) and Meylan \etal\ (1994, 1995) for the case of \cent.

\subsection{ The model }

The  ``lowered Maxwellian''  energy  dependence $[\exp(-A_iE)-1]$  has
been shown (King  1966) to be a good approximation  to the solution of
the  Fokker-Planck equation describing  the phase-space  diffusion and
evaporation  of  stellar systems  like  globular clusters.   Following
Eddington (1916), Michie (1963) introduced possible radial ($\overline
{v_r^2}$    $\not=$    $\overline    {v_{\theta}^2}$   =    $\overline
{v_{\phi}^2}$)) anisotropy by a factor of the form $\exp(-\beta J^2)$,
where $J$ is  the total angular momentum.  Da  Costa \& Freeman (1976)
showed that single-mass,  isotropic King models are unable  to fit the
entire  surface brightness  profile  of most  globular clusters;  they
generalized these  simple models to produce  more realistic multi-mass
models with full equipartition of energy in the centre.

In  the  present work,  we  use  a  multimass anisotropic  King-Michie
dynamical  model,  based  on  an  assumed  form  for  the  phase-space
distribution  function in  an  approach  similar to  that  of Gunn  \&
Griffin  (1979).   Each  of  the  twelve subpopulations  used  has  an
energy-angular momentum distribution  function $f_i(E,J)$ such that: %
$$ f_i(E,J)  \propto [\exp(-A_iE)-1]  \exp(-\beta J^2). \eqno(4)  $$ %
Thermal equilibrium is  assumed in the cluster center,  because of the
short  expected  relaxation  time,  in  order to  force  $A_i$  to  be
proportional to  the mean  mass $\overline{m}_i$ of  the stars  in the
subpopulation  considered. A  model is  specified by  an  initial mass
function (IMF)  exponent $x$ (see  below) and by four  parameters: (i)
the core  radius \rc, (ii) the  scale velocity \vs,  (iii) the central
value  of the  gravitational potential  \wo, and  (iv)  the anisotropy
radius  $r_a$, beyond  which  the velocity  dispersion tensor  becomes
increasingly radial.

\subsection{ The initial mass function }

There is no observational    determination  of the  present-day   mass
function in G1.  Consequently,  and in order  to mimic a real cluster,
main sequence (MS) stars, white dwarfs (wd) and heavier remnants (hr),
such  as  neutron stars, are  distributed   into twelve different mass
classes,   adopting the usual power-law  form   for  the initial  mass
function:
$$
dN \propto m^{-{x}} d\thinspace \rm{log}(m) \eqno(5)
$$
where  the exponent  $x$  would equal  1.35  in the  case of  Salpeter
(1955). 
 
This initial mass  function must be cut off  at both extremities.  The
upper limit has neither dynamical nor photometric influence because of
the rather  small initial  number of massive  stars which have  in any
case already evolved into dark stellar remnants.  This upper cutoff is
chosen rather arbitrarily at 100~\msun.   The lower limit is much more
controversial  because  of   the  potential  dynamical  importance  of
numerous  low-mass stars  with low-luminosity.   As found  by  Gunn \&
Griffin (1979), this lower mass cutoff,  if it is low enough, does not
significantly  affect the  cluster structure  as traced  by  the giant
stars.  Numerous  light stars  do not change  the quality of  the fit,
they simply increase the  cluster mass and mass-luminosity ratio.  The
individual mass of the lightest stars is taken equal to 0.13~\msun.

Owing   to  the total   lack    of observational  constraints on   the
present-day mass function along the main sequence, the exponent $x$ of
the initial mass function is  allowed to take  different values in the
following three mass intervals: $x_{hr}$, describing the heavy stellar
remnants, resulting from the already evolved stars with initial masses
in the  mass range   between    0.88 and 100~\msun;     $x_{MS}^{up}$,
describing the  stars still on the  main sequence, with initial masses
in  the  mass range between  0.32 and   0.88 \msun; and $x_{MS}^{low}$
describing the stars still  on the main  sequence, with initial masses
in the mass range between 0.13 and 0.32\msun.

\subsection{ The fit }
 
Both  the HST/WFPC2  images providing  the surface  brightness profile
(Table~3  and Fig.~5),  and  the integrated  light KECK/HIRES  spectra
providing the central velocity dispersion (Djorgovski \etal\ 1997) are
heavily dominated by  the light emitted by the  brightest stars in G1.
All of these stars, viz., the  giants and subgiants visible in the CMD
(see Fig.~2), have individual masses equal to or slightly smaller than
the  turnoff mass,  and belong  to the  same  subpopulation containing
stars  with  individual  masses  in  the  range  0.63  to  0.88~\msun.
Consequently,  the fits between  the models  and the  observations are
made by  using only this  subpopulation, which contains  the brightest
members in  the cluster, i.e.,  the giants, subgiants,  turnoff stars,
and the  stars at the top  of the main sequence.   An acceptable model
must  match, simultaneously,  first, the  observed  surface brightness
profile, fitted by least squares  to the integrated density profile of
the subpopulation  containing the bright  stars, as determined  by the
model (see  Fig.~5), and, second,  the observed value of  the velocity
dispersion in  the core  of the cluster,  which is  unfortunately less
constraining than the full  velocity dispersion profile, as available,
e.g., in the case of \cent\ (Meylan \etal\ 1995).
%
%
In addition to the above two  requirements, a model has to recover the
total luminosity of  the cluster, viz., \Mv\ =  --10.86~mag, to within
0.1~mag in order to be considered as satisfying.

\subsection{ Relaxation time }
 
Two  different  ``average'' relaxation  times  are  obtained for  each
model: a half-mass relaxation time  and a central relaxation time. The
term ``average''  means that  these times depend  on the  mean stellar
mass  of  the system,  instead  of  being  related to  one  particular
species. Spitzer \& Hart's (1971) standard formula transforms into:
$$
t_{rh} 
= (8.92~10^5 yr) 
\times
{{({\mathcal{M}}/{1m}{_\odot})^{1\over 2}} 
\over 
{({\overline {\mathcal{M}}}/{1m{_\odot}})}}
\times
{{({r_h/1pc})^{3\over 2}} 
\over 
{\rm {log}}({0.4 {\mathcal{M}}}/{\overline {m}})}
\eqno(6)
$$
where $\overline m$ is the mean  stellar mass of  all the stars in the
cluster, ${\mathcal{M}}$ the total mass of  the cluster, and $r_h$ the
half-mass radius.   From Lightman    \& Shapiro (1978),    the central
relaxation time is given by:
$$
t_{r,\circ} = (1.55~10^7 yr) 
\times
{{(1{m}_\odot/ \overline {{m}}_\circ)} 
\over
{\rm {log}}(0.5 {\mathcal{M}}/\overline {m})} 
\times
{{v_s} \over {1km/s}} 
\times
({{r_c} \over {1pc}})^2
\eqno(7)
$$
where $\overline {m}_\circ$ is  the mean mass of  all the particles in
thermal equilibrium in  the central parts,  $v_s$ the  velocity scale,
and $r_c$ the core radius.
         
\subsection{ Results from King-Michie models }

An  extensive grid of about  150,000  models is computed  in order  to
explore the parameter space defined by the Initial Mass Function (IMF)
exponents  $x$  (defined over  three  different mass  ranges $x_{hr}$,
$x_{MS}^{up}$, and $x_{MS}^{low}$, and  where $x$ would equal  1.35 in
the case  of  Salpeter 1955),   the  central gravitational   potential
$W_{\circ}$, and the anisotropy radius \ra.

Good models are considered as such not only on the basis of the \chis\
of the surface  brightness fit --- the topology  of the \chis\ surface
has no unique  minimum --- but also on the  basis of their predictions
of  the integrated luminosity  and mass-to-light  of the  cluster.  We
present hereafter results for the twelve models with lowest chi-square
and fulfilling the other two requirements.

The different columns in  Table~4 give, for each  model defined by its
index, the central  value of the gravitational  potential \wo; its  MS
mass function exponents $x_{MS}^{up}$ and $x_{MS}^{low}$; the fraction
\Mhr\  of its  total mass  in  the form   of stellar remnants  such as
neutron stars  and white dwarfs;  its concentration \conc, core radius
\rc, half-mass radius \rh, and tidal  radius \rt; its central value of
the mass density \rhoc, mean  mass density \rhoh\ inside the half-mass
radius, and mean mass density \rhot\ inside the tidal radius.

The different columns  in Table~5 give, for each  model defined by its
index,  the total mass  \Mtot\ of  the cluster,  in millions  of solar
masses, and its corresponding  mass-to-light ratio \mlv, also in solar
units; the  half-mass relaxation time \trh\ from  Eq.~(6), and central
relaxation time \tro\ from Eq.~(7).

Since the velocity  dispersion profile is reduced to  one single value
--- the central velocity dispersion --- the models are not constrained
as strongly as in the case of \cent\ (Meylan \etal\ 1995), and equally
good  fits  are obtained  for  rather  different  sets of  parameters.
Consequently, reliable results only  relate to general parameters like
the concentration  and the total mass,  but probably fail  in any more
detailed parameters.  Nevertheless, very  large areas of the parameter
space can be eliminated with  confidence since they never generate any
successful models.

The  IMF exponent  $x_{hr}$, describing  the amount  of  heavy stellar
remnants, appears  in all good models to  be very close to  $x$ = 1.35
(Salpeter 1955). The IMF  exponent $x_{MS}^{up}$, describing the upper
part of the MS,  appears in all good models to be  very close to $x$ =
1.55.  The IMF  exponent $x_{MS}^{low}$, describing the  lower part of
the MS, is  not very well constrained; this is  in agreement with Gunn
\&  Griffin's  (1979)  findings  that  the  lower-mass  stars  do  not
significantly  affect the  cluster structure  as traced  by  the giant
stars.  The fraction  of the total mass of the cluster  in the form of
heavy stellar remnants  (neutron stars and white dwarfs)  is always in
the range of 18 to 20\%.  

With a concentration  \conc\ somewhere around 2.5, G1  is definitely a
very concentrated  globular cluster:  all good models  present clearly
all the characteristics  of a collapsed cluster.  With  its small core
radius, G1 is hardly resolved  in its central parts while its envelope
exhibits an extended profile typical  of a collapsed cluster (see also
Djorgovski 1988).
%
%
The core radius  has a mean value of about  \rc\ $\simeq$ 0.52~pc, the
half-mass radius \rh\ $\simeq$ 13.5~pc, and the tidal radius, the most
uncertain  of  these three  radii,  has a  mean  value  of about  \rt\
$\simeq$ 200~pc.  The corresponding mass densities have mean values of
about  \rhoc\ =  4.7 $\times$  10$^5$  \msun pc$^{-3}$,  \rhoh\ =  7.5
$\times$ 10$^2$ \msun pc$^{-3}$, \rhot\ = 4.2 $\times$ 10$^{-1}$ \msun
pc$^{-3}$.

With a  total mass  somewhere between  14 and 17  \milm, and  with the
corresponding  mass-to-light ratio \mlv\  between 6.9  and 8.1,  G1 is
significantly  more massive than  \cent, maybe  by up  to a  factor of
three.  These  King-Michie mass estimates  are in full  agreement with
the  King mass  estimate (\Mass$_K$  = 15  $\times$ \milm\  with \mlv\
$\simeq$  7.5), while the  Virial mass  estimate (\Mass$_{Vir}$  = 7.3
$\times$ \milm\ with \mlv\ $\simeq$  3.6) is smaller by about a factor
of two.   It is worth mentioning  that such a  difference between King
and Virial  mass estimates is not  specific to G1: the  same factor of
about two  is also  observed between the  King-Michie and  Virial mass
estimates of any cluster.  See, e.g., in Table~6 the comparison of the
results obtained for G1 and \cent\ (Meylan \& Mayor 1986, Meylan 1987,
Meylan \etal\ 1995, and this paper).

\section{ Is Mayall~II $\equiv$~G1 a genuine globular cluster~? }

We reach  the following conclusions: {\bf  (i)} G1 is  only the second
globular cluster in which  we observe a significant metallicity spread
among its  giant stars, \cent\ being  the first such  case; {\bf (ii)}
All mass estimates (Table~6) give a total mass for G1 equal to as much
as  three times  the  total mass  of  \cent; {\bf  (iii)} With  \conc\
$\simeq$ 2.5, G1 is  significantly more concentrated than \tuca, which
is a massive Galactic globular cluster considered on the verge of core
collapse;  all  G1  structural  parameters deduced  from  its  surface
brightness profile  as well  as the model  densities are typical  of a
collapsed  cluster; {\bf  (iv)}  G1  is the  heaviest  of the  weighed
globular clusters.

Although  \cent\ is  by far  the brightest  and most  massive globular
cluster in  our Galaxy, G1 may  not be the only  such massive globular
cluster belonging to M31.  This galaxy, which has about three times as
many globular clusters as our  Galaxy, has at least three other bright
clusters which have central  velocity dispersions larger than 20 \kms\
(Djorgovski \etal\ 1997).  Unfortunately, so  far, G1 is the only such
cluster  imaged with  the  high spatial  resolution  of the  HST/WFPC2
camera, and  consequently the  only such massive  cluster with  a high
quality  surface brightness profile  and known  structural parameters.
G1 and the other three bright M31 globular clusters represent probably
the  high-mass and high-luminosity  tails of  the mass  and luminosity
distributions of the rich population of M31 globular clusters.

The above large mass estimates, implying a rather deep potential well,
obviously  relate to  the  metallicity spread  whose  origin is  still
unknown.  There  are essentially three different  scenarios to explain
such  a metallicity  spread:  (i) metallicity  self-enrichment in  the
globular  cluster, (ii)  primordial metallicity  inhomogeneity  in the
proto-cluster cloud, and (iii)  the present globular cluster is merely
the remaining core of a previously larger entity.

Even more so than \cent, G1 could be a kind of transition step between
globular  clusters  and  dwarf   elliptical  galaxies,  in  being  the
remaining  core of  a  dwarf  galaxy whose  envelope  would have  been
severely pruned  by tidal shocking  due to the  bulge and disk  of its
host galaxy, M31.

Kormendy (1985, 1987, 1994) used the four following quantities --- the
central  surface  brightness \muov,  the  central velocity  dispersion
\sigpo, the  core radius \rc,  and the total absolute  magnitude M$_V$
--- in order to define various  planes from combinations of two of the
above  four parameters,  e.g.,  (\muov\ vs.   log~\rc).   In all  four
planes  plotted by Kormendy  (1985, his  Fig.~3), the  various stellar
systems segregate into three well separated sequences: (i) ellipticals
and bulges, (ii) dwarf ellipticals, and (iii) globular clusters.  When
plotted on any  of the four planes, G1 appears  always on the sequence
of  globular clusters,  and  cannot be  confused  or assimilated  with
either ellipticals and bulges or  dwarf ellipticals.  The same is true
for \cent.  Consequently, in spite  of their large masses and internal
stellar metallicity  spreads, G1 and  \cent\ look like  genuine bright
and massive globular clusters.  But  where on these diagrams would the
remaining cores of dwarf galaxies be located~?

Because of its  very large central velocity dispersion,  M32 could not
be the  progenitor of  a star  cluster like G1  (van der  Marel \etal\
1998).  But this  is not true for the  nucleated dwarf galaxy NGC~205,
which  has a  central velocity  dispersion  similar to  G1 (Carter  \&
Sadler 1990,  Held \etal\ 1990,  Bender \etal\ 1991).  The  nucleus of
NGC~205 is  the only one for  which the values of  the four parameters
used  in Kormendy's  diagrams are  known, viz.,  the  central velocity
dispersion  \sigpo\  = 30  \kms\  (Bender  \etal\  1991), the  central
surface brightness  \muov\ = 12.84~mag~arcsec$^{-2}$,  the core radius
\rc\ = 0.35~pc, and the  total absolute magnitude M$_V$ = --9.6 (Jones
\etal\ 1996).   These values place NGC~205 nucleus,  in the Kormendy's
diagrams,  very close to  G1, right  on the  sequence of  the globular
clusters.  Although  this does not  prove that all  (massive) globular
clusters are the  remnant cores of nucleated dwarf  galaxies, it shows
that at least  the nucleus of one such  dwarf exhibits characteristics
identical to those  of globular clusters.  It would  be useful to know
more about the nuclei of other dwarf galaxies.

Of  these  massive  globular  clusters,  the most  nearby,  \cent,  is
naturally the best studied, nevertheless without decreasing the number
of its  conundrums.  For instance, recent  photometric (Anderson 1997)
and kinematic  (Norris \etal\ 1997) studies show  that \cent\ presents
numerous  characteristics about its  stellar populations  which remain
without any explanation, if  not completely puzzling.  Presently, only
the anti-correlation  of metallicity  with age (Hughes  \& Wallerstein
2000) and  the unusual  patterns of  CN elements  (Hilker  \& Richtler
2000) recently observed  in \cent\ suggest< that  this cluster enriched
itself over a timescale of  about 3~Gyr.  This contradicts the general
idea that  all the  stars in  a globular cluster  are coeval,  and may
favor the  origin of \cent\  as being the  remaining core of  a larger
entity, e.g., of  a former nucleated dwarf elliptical  galaxy.  Such a
general  idea  had already  been  suggested  by  Zinnecker (1988)  and
Freeman (1993).  In  any case, the very massive  globular clusters, by
the mere  fact that their  large masses imply complicated  stellar and
dynamical  evolution,  may  blur  the  former  clear  (or  simplistic)
difference between globular clusters and dwarf galaxies.


\acknowledgements

AS was supported by  the National Aeronautics and Space Administration
(NASA)  grants HF-01077.01-94A,  GO-05907.01-94A,  and GO-06477.02-95A
from the Space  Telescope Science Institute, which is  operated by the
Association  of Universities  for Research  in Astronomy,  Inc., under
NASA  contract NAS5-26555.   SGD was  supported in  part by  the grant
GO-060399 from STScI.  GM thanks  Jennifer Lotz (JHU), Roeland van der
Marel (STScI),  and Brad Whitmore (STScI)  for interesting discussions
and information about NGC 205.





\medskip   

%
\begin{deluxetable}{l r} 
\tablecaption{ General information about Mayall~II $\equiv$~G1 }
\footnotesize
\tablewidth{0pt}
\tablehead{
        \colhead{Parameters} &
        \colhead{Mayall~II $\equiv$~G1} 
}
\startdata
$\alpha$ G1 (J2000)             &  00$^h$  32$^m$  46.6$^s$    \\
$\delta$ G1 (J2000)             &  +39$^o$  34\arcm\ 40\arcs\  \\
$\alpha$ M31 (J2000)            &  00$^h$  42$^m$  44.5$^s$    \\
$\delta$ M31 (J2000)            &  +41$^o$  16\arcm\ 29\arcs\  \\
  Distance $D$ to M31           &  770 kpc                     \\
  Color excess $E(B-V)$         &  0.06 mag                    \\
  True distance modulus $(m-M)$ &  24.42 mag                   \\
  Observed magnitude $V$        &  13.48 mag                   \\
  Absolute magnitude $M_V$      & --10.94 mag                  \\
  Central $V$ surf bright \muov\ &  13.47 mag arcsec$^{-2}$    \\
  Age                           &  $\sim$ 15 Gyr               \\
  Metallicity [Fe/H]            & --0.95                       \\
  Mean ellipticity $\epsilon$   &   0.2                        \\
  Radial velocity $V_r$         & --331 \pmm\ 24 \kms          \\   
  Velocity dispersion \sobs\    &  25.1 \kms                   \\
  Vel. dis. aperture corrected $\sigma$(0) &  27.8 \kms        \\ 
\enddata
\normalsize
\end{deluxetable}


%

%

%
\begin{deluxetable}{ c c c } 
\tablecaption{ Mayall~II $\equiv$~G1: ellipticity $\epsilon$  and 
               position angle $PA$ as a function of the semi-major axis $a$ }
\footnotesize
\tablewidth{0pt}
\tablehead{
        \colhead{$a$} &
        \colhead{$\epsilon$} &
        \colhead{$PA$} \\
        \colhead{[arcsec]} &
        \colhead{} &
        \colhead{[degree]} 
}
\startdata
   0.091 &   0.139 $\pm$ 0.010 &  109.2 $\pm$  1.0 \\
   0.137 &   0.151 $\pm$ 0.010 &  109.2 $\pm$  1.0 \\
   0.182 &   0.087 $\pm$ 0.010 &  105.3 $\pm$  1.0 \\
   0.228 &   0.113 $\pm$ 0.010 &  110.0 $\pm$  1.0 \\
   0.273 &   0.112 $\pm$ 0.010 &  114.4 $\pm$  1.0 \\
   0.319 &   0.110 $\pm$ 0.010 &  120.6 $\pm$  1.0 \\
   0.364 &   0.115 $\pm$ 0.010 &  122.5 $\pm$  1.0 \\
   0.410 &   0.120 $\pm$ 0.010 &  123.7 $\pm$  1.0 \\
   0.455 &   0.123 $\pm$ 0.010 &  124.0 $\pm$  1.0 \\
   0.501 &   0.127 $\pm$ 0.010 &  123.3 $\pm$  1.0 \\
   0.546 &   0.133 $\pm$ 0.010 &  123.5 $\pm$  1.0 \\
   0.591 &   0.149 $\pm$ 0.010 &  123.6 $\pm$  1.0 \\
   0.637 &   0.178 $\pm$ 0.010 &  123.5 $\pm$  1.0 \\
   0.683 &   0.192 $\pm$ 0.010 &  123.4 $\pm$  1.0 \\
   0.728 &   0.192 $\pm$ 0.010 &  123.4 $\pm$  1.0 \\
   0.774 &   0.192 $\pm$ 0.010 &  123.5 $\pm$  1.0 \\
   0.819 &   0.193 $\pm$ 0.010 &  122.4 $\pm$  1.0 \\
   0.865 &   0.193 $\pm$ 0.010 &  120.5 $\pm$  1.0 \\
   0.910 &   0.193 $\pm$ 0.010 &  119.8 $\pm$  1.0 \\
   0.956 &   0.195 $\pm$ 0.010 &  119.8 $\pm$  1.0 \\
   1.046 &   0.227 $\pm$ 0.029 &  119.1 $\pm$  1.6 \\
   1.183 &   0.235 $\pm$ 0.011 &  119.1 $\pm$  1.7 \\
   1.342 &   0.199 $\pm$ 0.024 &  117.4 $\pm$  2.3 \\
   1.501 &   0.193 $\pm$ 0.015 &  116.2 $\pm$  2.9 \\
   1.683 &   0.242 $\pm$ 0.030 &  120.5 $\pm$  5.9 \\
   1.888 &   0.269 $\pm$ 0.017 &  126.1 $\pm$  1.3 \\
   2.115 &   0.299 $\pm$ 0.020 &  124.4 $\pm$  2.0 \\
   2.387 &   0.251 $\pm$ 0.031 &  122.8 $\pm$  2.5 \\
   2.661 &   0.250 $\pm$ 0.029 &  124.0 $\pm$  3.2 \\
   2.979 &   0.231 $\pm$ 0.018 &  123.5 $\pm$  1.9 \\
   3.564 &   0.216 $\pm$ 0.021 &  122.0 $\pm$ 11.9 \\
   4.495 &   0.234 $\pm$ 0.023 &  123.0 $\pm$ 11.5 \\
   5.653 &   0.208 $\pm$ 0.032 &  121.0 $\pm$ 10.0 \\
   7.105 &   0.254 $\pm$ 0.062 &  120.6 $\pm$  9.0 \\
   8.944 &   0.183 $\pm$ 0.026 &  123.5 $\pm$  8.4 \\
  11.259 &   0.146 $\pm$ 0.085 &  129.3 $\pm$  6.1 \\
\enddata
\normalsize
\end{deluxetable}


%
%
%

%
\begin{deluxetable}{r c c l r c c} 
\tablecaption{ Mayall~II $\equiv$~G1: surface brightness $\mu_V$ 
               and integrated $V$ magnitude as a function of the radius }
\footnotesize
\tablewidth{0pt}
\tablehead{
        \colhead{$R$} &
        \colhead{$\mu_V$} &
        \colhead{$V_{int}$} &
        \colhead{   } &
        \colhead{$R$} &
        \colhead{$\mu_V$} &
        \colhead{$V_{int}$}  \\
        \colhead{[arcsec]} &
        \colhead{[mag]} &
        \colhead{[mag]} &
        \colhead{   } &
        \colhead{[arcsec]} &
        \colhead{[mag]} &
        \colhead{[mag]} 
}
\startdata
 0.039 & 13.467 $\pm$ 0.019 & 19.268 & &  1.243 & 17.307 $\pm$ 0.055 & 14.206 \\
 0.061 & 13.557 $\pm$ 0.022 & 18.321 & &  1.251 & 17.329 $\pm$ 0.055 & 14.202 \\
 0.066 & 13.574 $\pm$ 0.014 & 18.161 & &  1.512 & 17.855 $\pm$ 0.094 & 14.096 \\
 0.074 & 13.610 $\pm$ 0.015 & 17.930 & &  1.586 & 17.953 $\pm$ 0.099 & 14.073 \\
 0.085 & 13.657 $\pm$ 0.026 & 17.655 & &  1.608 & 17.963 $\pm$ 0.095 & 14.066 \\
 0.089 & 13.689 $\pm$ 0.032 & 17.566 & &  1.826 & 18.241 $\pm$ 0.091 & 14.005 \\ 
 0.100 & 13.745 $\pm$ 0.019 & 17.345 & &  2.025 & 18.483 $\pm$ 0.085 & 13.958 \\
 0.108 & 13.747 $\pm$ 0.020 & 17.198 & &  2.189 & 18.673 $\pm$ 0.083 & 13.925 \\
 0.130 & 13.877 $\pm$ 0.014 & 16.854 & &  2.206 & 18.722 $\pm$ 0.087 & 13.922 \\
 0.138 & 13.896 $\pm$ 0.017 & 16.748 & &  2.585 & 19.107 $\pm$ 0.094 & 13.861 \\
 0.157 & 14.026 $\pm$ 0.023 & 16.523 & &  2.665 & 19.191 $\pm$ 0.099 & 13.850 \\
 0.176 & 14.093 $\pm$ 0.016 & 16.332 & &  2.980 & 19.497 $\pm$ 0.097 & 13.812 \\
 0.186 & 14.144 $\pm$ 0.014 & 16.241 & &  3.219 & 19.746 $\pm$ 0.094 & 13.788 \\
 0.189 & 14.163 $\pm$ 0.012 & 16.215 & &  3.299 & 19.795 $\pm$ 0.099 & 13.781 \\
 0.225 & 14.329 $\pm$ 0.014 & 15.940 & &  3.889 & 20.224 $\pm$ 0.114 & 13.734 \\
 0.228 & 14.357 $\pm$ 0.014 & 15.921 & &  4.056 & 20.322 $\pm$ 0.116 & 13.723 \\
 0.253 & 14.472 $\pm$ 0.017 & 15.767 & &  4.211 & 20.447 $\pm$ 0.114 & 13.714 \\
 0.276 & 14.574 $\pm$ 0.020 & 15.644 & &  4.697 & 20.715 $\pm$ 0.119 & 13.687 \\
 0.287 & 14.634 $\pm$ 0.025 & 15.590 & &  5.377 & 21.011 $\pm$ 0.117 & 13.656 \\
 0.333 & 14.857 $\pm$ 0.027 & 15.396 & &  5.520 & 21.074 $\pm$ 0.117 & 13.650 \\
 0.344 & 14.882 $\pm$ 0.027 & 15.355 & &  5.675 & 21.175 $\pm$ 0.117 & 13.644 \\
 0.367 & 14.988 $\pm$ 0.030 & 15.276 & &  6.855 & 21.721 $\pm$ 0.141 & 13.605 \\
 0.403 & 15.129 $\pm$ 0.029 & 15.166 & &  6.863 & 21.692 $\pm$ 0.135 & 13.605 \\
 0.468 & 15.354 $\pm$ 0.027 & 14.999 & &  7.514 & 21.944 $\pm$ 0.148 & 13.589 \\
 0.469 & 15.354 $\pm$ 0.029 & 14.996 & &  8.281 & 22.242 $\pm$ 0.175 & 13.572 \\
 0.486 & 15.423 $\pm$ 0.032 & 14.958 & &  8.758 & 22.380 $\pm$ 0.187 & 13.563 \\
 0.588 & 15.786 $\pm$ 0.050 & 14.767 & & 10.002 & 22.768 $\pm$ 0.226 & 13.543 \\
 0.598 & 15.808 $\pm$ 0.050 & 14.752 & & 10.226 & 22.784 $\pm$ 0.225 & 13.540 \\ 
 0.638 & 15.926 $\pm$ 0.051 & 14.693 & & 11.179 & 23.063 $\pm$ 0.266 & 13.528 \\ 
 0.710 & 16.133 $\pm$ 0.055 & 14.600 & & 12.084 & 23.283 $\pm$ 0.320 & 13.518 \\ 
 0.763 & 16.287 $\pm$ 0.058 & 14.541 & & 13.919 & 23.849 $\pm$ 0.458 & 13.502 \\ 
 0.858 & 16.580 $\pm$ 0.066 & 14.452 & & 14.269 & 23.991 $\pm$ 0.516 & 13.500 \\ 
 0.868 & 16.577 $\pm$ 0.064 & 14.444 & & 14.598 & 24.113 $\pm$ 0.556 & 13.498 \\ 
 0.974 & 16.836 $\pm$ 0.065 & 14.363 & & 17.632 & 24.941 $\pm$ 0.987 & 13.484 \\ 
 1.036 & 16.940 $\pm$ 0.064 & 14.322 & & 18.214 & 25.051 $\pm$ 1.054 & 13.482 \\ 
 1.182 & 17.188 $\pm$ 0.057 & 14.237 & & 18.945 & 25.190 $\pm$ 1.137 & 13.480 \\ 
\enddata
\normalsize
\end{deluxetable}

%
%

\begin{deluxetable}{ c c c c c c c c c c c c c } 
\tablecaption{ Mayal II $\equiv$~G1: multi-mass King-Michie model parameters }
\footnotesize
\tablewidth{0truecm}
\tablehead{
        \colhead{model} &
        \colhead{\wo} &
        \colhead{$x_{MS}^{up}$} &
        \colhead{$x_{MS}^{low}$} &
        \colhead{$M_{hr+wd}$} &
        \colhead{$c$} &
        \colhead{\rc} &
        \colhead{\rh} &
        \colhead{\rt} &
        \colhead{\rhoo} &
        \colhead{\rhoh} &
        \colhead{\rhot} \\
        \colhead{index} &
        \colhead{} &
        \colhead{} &
        \colhead{} &
        \colhead{\%} &
        \colhead{log ($r_t/r_c$)} &
        \colhead{[pc]} &
        \colhead{[pc]} &
        \colhead{[pc]} &
        \colhead{[\msun pc$^{-3}$]} &
        \colhead{[\msun pc$^{-3}$]} &
        \colhead{[\msun pc$^{-3}$]} \\
        \colhead{(1)} &
        \colhead{(2)} &
        \colhead{(3)} &
        \colhead{(4)} &
        \colhead{(5)} &
        \colhead{(6)} &
        \colhead{(7)} &
        \colhead{(8)} &
        \colhead{(9)} &
        \colhead{(10)} &
        \colhead{(11)} &
        \colhead{(12)} 
}
\startdata
  1  & 15.25 & 1.50 & --0.50 & 20 & 2.49 & 0.53 & 12.3 & 163 & 
                                 4.7E+5 & 8.9E+2 & 7.8E--1   \\
  2  & 15.50 & 1.55 & --0.50 & 20 & 2.59 & 0.52 & 13.2 & 199 & 
                                 4.9E+5 & 7.7E+2 & 4.5E--1   \\
  3  & 15.50 & 1.55 & --0.40 & 19 & 2.57 & 0.52 & 13.2 & 193 & 
                                 4.8E+5 & 7.7E+2 & 4.9E--1   \\
  4  & 15.50 & 1.55 & --0.30 & 19 & 2.55 & 0.53 & 13.2 & 187 & 
                                 4.7E+5 & 7.8E+2 & 5.4E--1   \\
  5  & 15.75 & 1.55 &   0.00 & 18 & 2.61 & 0.52 & 13.9 & 212 & 
                                 4.8E+5 & 6.9E+2 & 3.9E--1   \\
  6  & 15.50 & 1.60 & --0.50 & 19 & 2.56 & 0.53 & 13.2 & 193 & 
                                 4.7E+5 & 7.7E+2 & 4.9E--1   \\
  7  & 15.50 & 1.60 & --0.40 & 19 & 2.55 & 0.53 & 13.2 & 187 & 
                                 4.6E+5 & 7.7E+2 & 5.4E--1   \\
  8  & 15.75 & 1.60 & --0.30 & 19 & 2.65 & 0.51 & 14.1 & 230 & 
                                 4.9E+5 & 6.7E+2 & 3.1E--1   \\
  9  & 15.75 & 1.60 & --0.20 & 18 & 2.63 & 0.52 & 14.0 & 221 & 
                                 4.8E+5 & 6.8E+2 & 3.5E--1   \\
 10  & 15.75 & 1.60 & --0.10 & 18 & 2.61 & 0.52 & 14.0 & 213 & 
                                 4.7E+5 & 6.8E+2 & 3.9E--1   \\
 11  & 15.75 & 1.60 &   0.00 & 18 & 2.59 & 0.53 & 14.0 & 205 & 
                                 4.6E+5 & 6.9E+2 & 4.4E--1   \\
 12  & 16.00 & 1.60 &  +0.20 & 17 & 2.68 & 0.52 & 15.0 & 245 & 
                                 4.8E+5 & 5.9E+2 & 2.6E--1   \\
\enddata
\normalsize
\end{deluxetable}


\begin{deluxetable}{ c c c c c } 
\tablecaption{ Mayal II $\equiv$~G1: multi-mass King-Michie model parameters }
\footnotesize
\tablewidth{0pt}
\tablehead{
        \colhead{model} &
        \colhead{\Mtot} &
        \colhead{\mlv} &
        \colhead{\trh} &
        \colhead{\tro} \\
        \colhead{index} &
        \colhead{[\milm]} &
        \colhead{[$\odot$]} &
        \colhead{[\bily]} &
        \colhead{[\mily]} \\
        \colhead{(1)} &
        \colhead{(2)} &
        \colhead{(3)} &
        \colhead{(4)} &
        \colhead{(5)} 
}
\startdata
  1 & 14 & 6.9 & 43 & 15  \\
  2 & 15 & 7.1 & 50 & 14  \\
  3 & 15 & 7.2 & 50 & 15  \\
  4 & 15 & 7.3 & 50 & 15  \\
  5 & 16 & 7.6 & 57 & 14  \\
  6 & 15 & 7.3 & 51 & 15  \\
  7 & 15 & 7.3 & 51 & 15  \\
  8 & 16 & 7.5 & 57 & 14  \\
  9 & 16 & 7.6 & 57 & 15  \\
 10 & 16 & 7.7 & 58 & 15  \\
 11 & 16 & 7.8 & 58 & 15  \\
 12 & 17 & 8.1 & 67 & 15  \\
\enddata
\normalsize
\end{deluxetable}


\begin{deluxetable}{l c c c c } 
\tablecaption{ Three  different mass  estimates for the two globular clusters 
               Mayall~II $\equiv$~G1 and NGC~5139 $\equiv$ \cent }
\footnotesize
\tablewidth{0pt}
\tablehead{
        \colhead{ Mass } &
        \colhead{ G1 }  &
        \colhead{ G1 }  &
        \colhead{ \cen } &
        \colhead{ \cen } \\
        \colhead{ Estimator      } &
        \colhead{ mass [\milm]   } &
        \colhead{ \mlv [$\odot$] } &
        \colhead{ mass [\milm]   } &
        \colhead{ \mlv [$\odot$] }  
}
\startdata
King         &  15.  &  7.5    &  4.3  &  3.5  \\
Virial       &  7.3  &  3.6    &  3.0  &  2.4  \\
King-Michie  & 14-17 & 6.9-8.1 &  5.1  &  4.1  \\
\enddata   
\normalsize
\end{deluxetable}


%
\begin{figure}[h]
\centerline{
\psfig{bbllx=15mm,bblly=35mm,bburx=185mm,bbury=247mm,%
figure=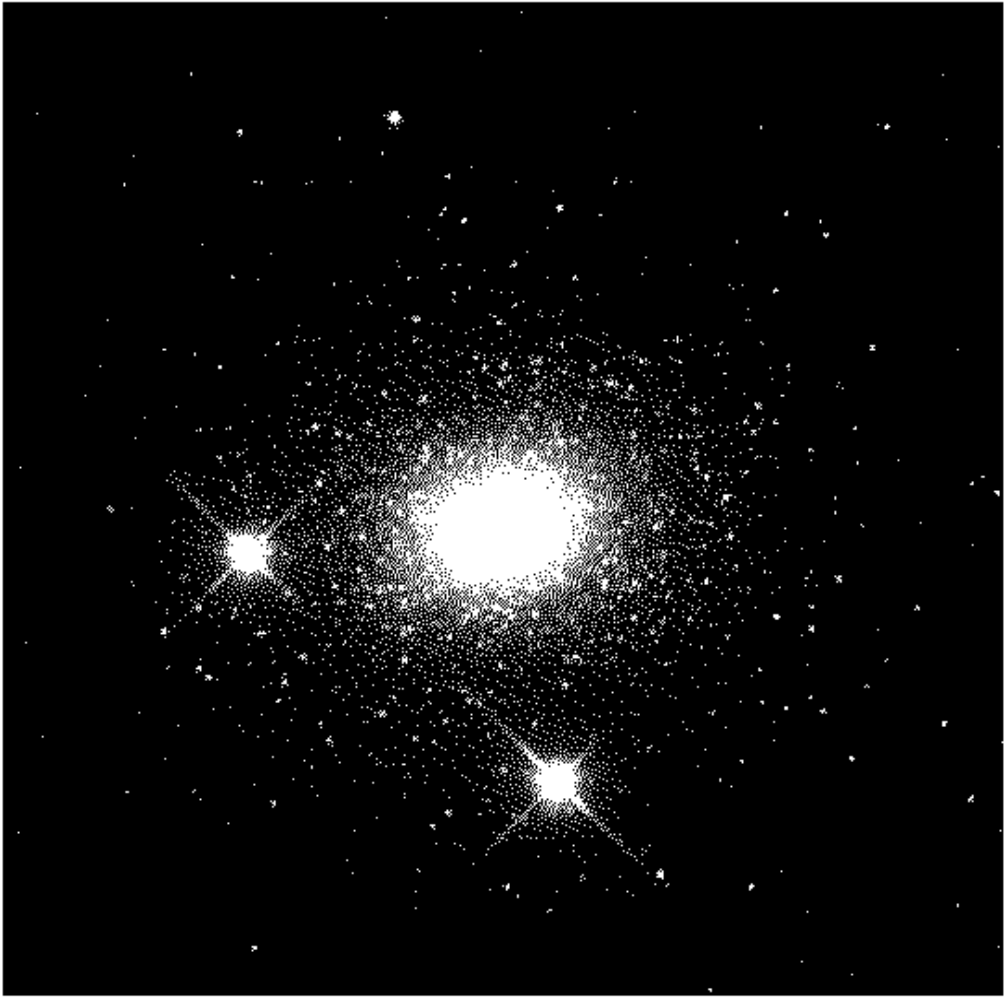,width=135mm}}
\caption[]{\label{fig.diagram} Mayall~II $\equiv$~G1  as seen with the
PC chip of  the WFPC2 camera onboard the  Hubble Space Telescope.  The
cluster is surrounded by two bright foreground stars.  This image is a
composite  from  F555W ($V$)  and  F814W  ($I$)  frames with  a  total
exposure time  of 2,200 seconds  in $V$ and  of 1,800 seconds  in $I$.
The  field  is  31.5\arcsec\   \x\  31.5\arcsec.   North  is  131\deg\
clockwise from  vertical direction and East is  41\deg\ clockwise from
vertical direction.  (A non-degraded color  version of this  figure is
available from the first author). }
\end{figure}


%
\begin{figure}[h]
\centerline{
\psfig{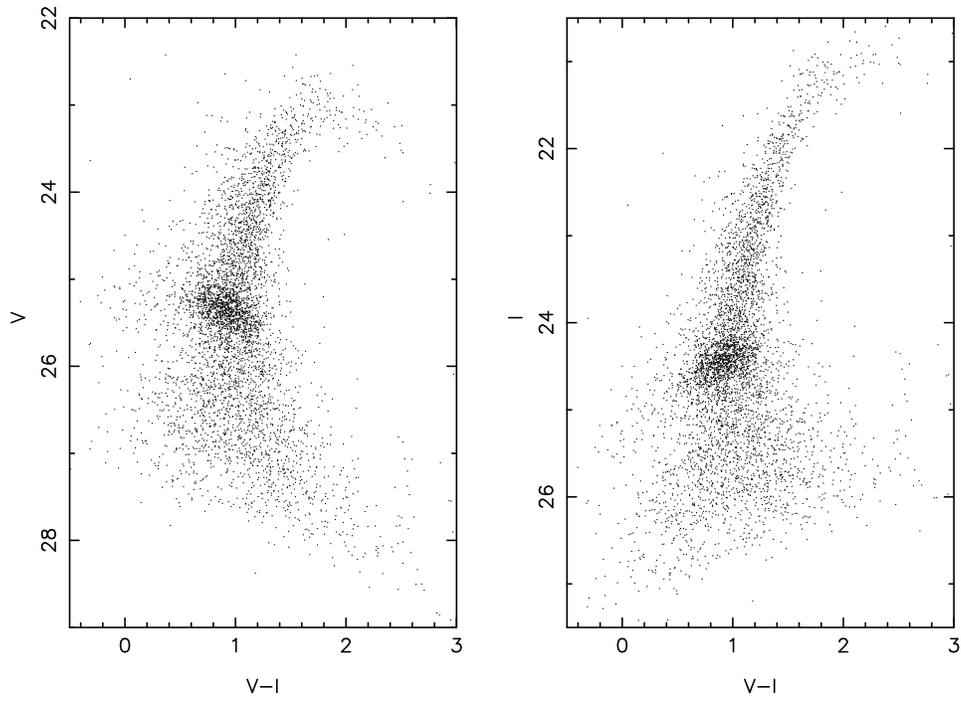}}
\caption[]{\label{fig.diagram}   Color-magnitude   diagram  (CMD)   of

Mayall~II $\equiv$~G1.   The left  panel shows the  $V$~vs.~$V-I$ CMD,
while  the right  panel  displays the  $I$~vs.~$V-I$  CMD. Each  panel
contains the same 4903 stars.
}
\end{figure}


%
\begin{figure}[h]
\centerline{
\psfig{bbllx=33mm,bblly=17mm,bburx=161mm,bbury=260mm,%
figure=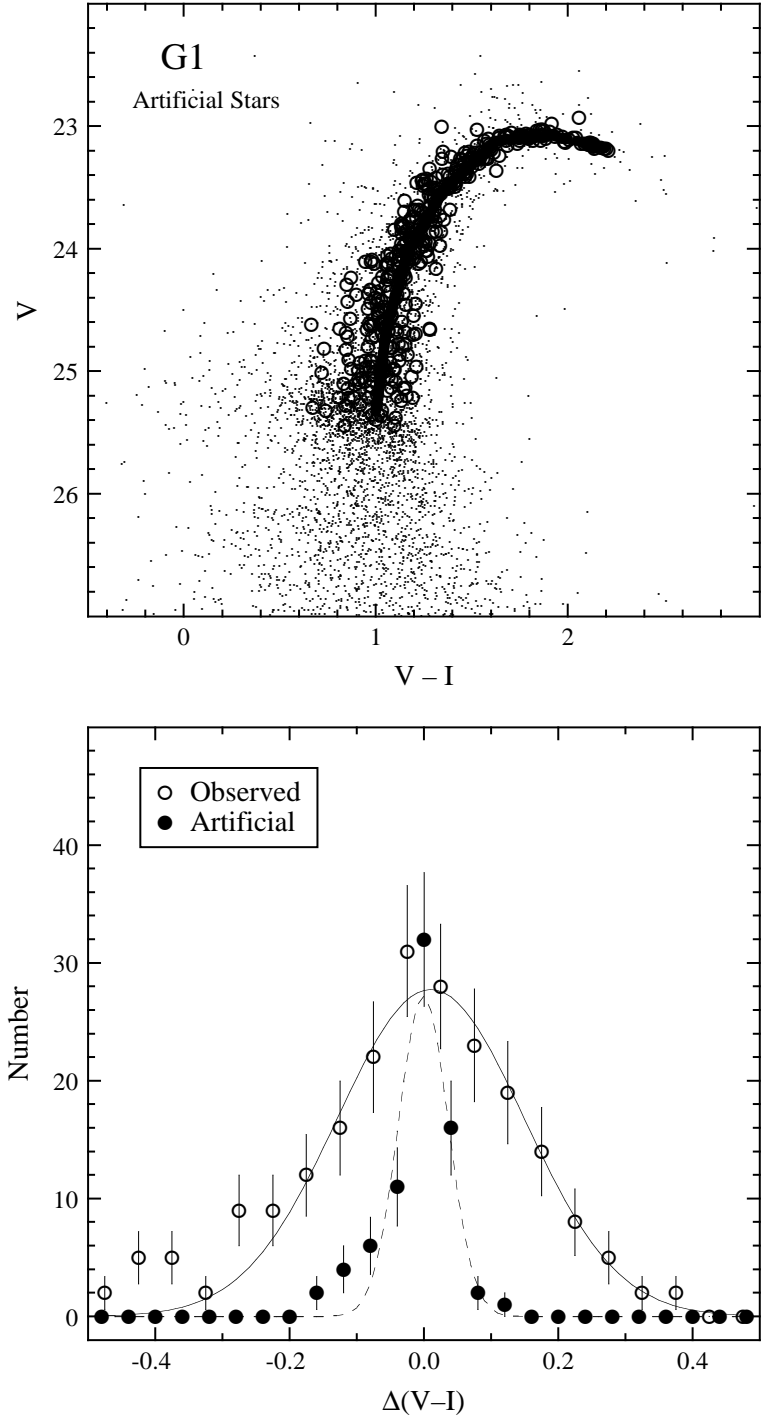,width=100mm}}
\caption[]{\label{fig.diagram}  Artificial  star  experiments.~  Upper
panel:  color-magnitude  diagram where  filled  circles represent  the
original magnitudes and  colors of the 420 artificial  stars while the
open  circles are  their  recovered values.   The  dots represent  the
genuine G1  stars.  Lower  panel: the color  widths of  artificial and
observed stars.   The open circles  show the color histogram  of stars
located 1.8  \pmm\ 0.25 magnitudes  above the horizontal branch  of G1
while filled circles show the color histogram of the artificial stars.
}
\end{figure}


%
\begin{figure}[h]
\centerline{
\psfig{bbllx=15mm,bblly=35mm,bburx=185mm,bbury=247mm,%
figure=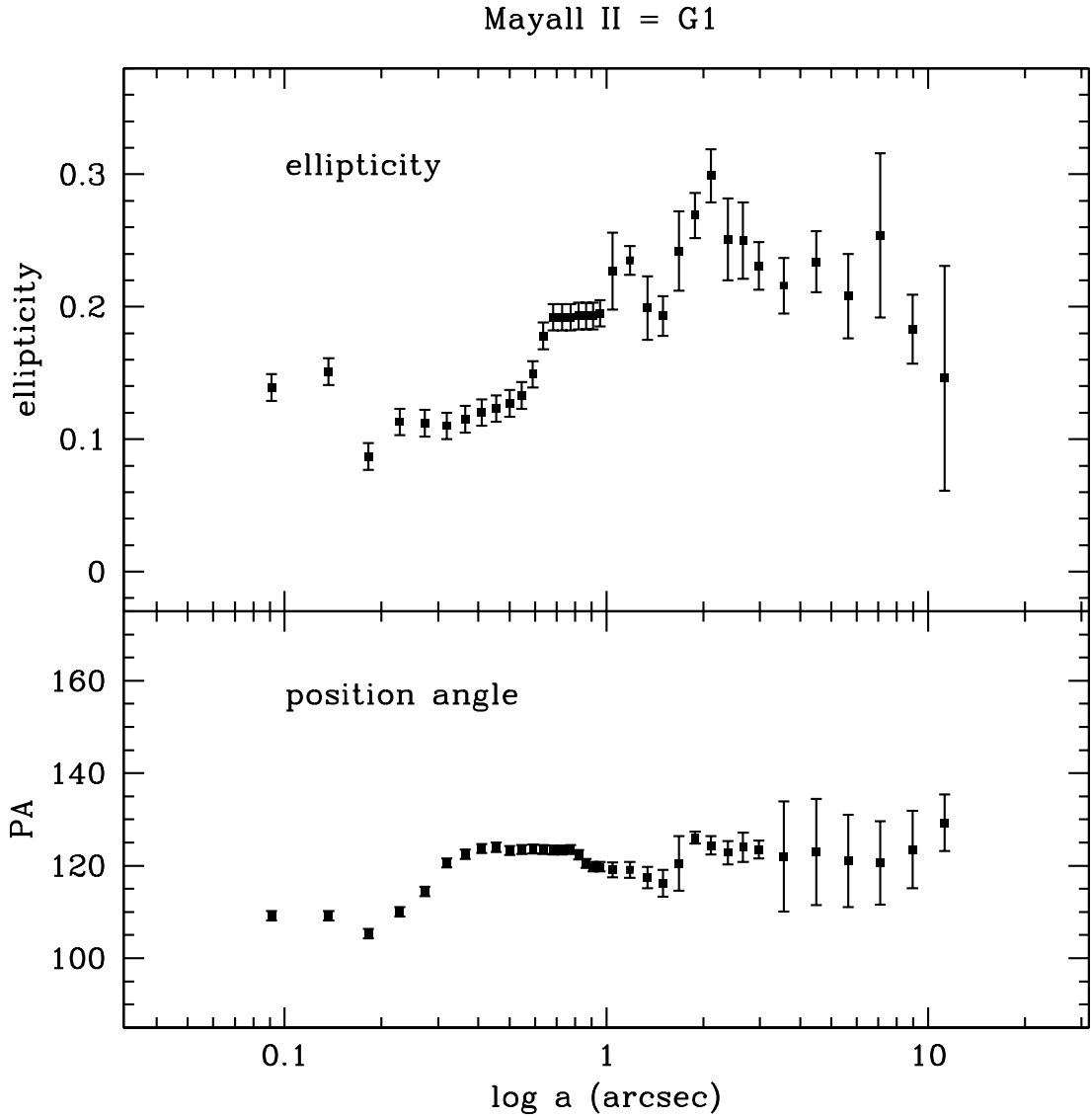,width=135mm}}
\caption[]{\label{fig.diagram}  Ellipticity and  position   angle as a
function of the semi-major axis, for  Mayall~II $\equiv$~G1, using the
stack of all F555W ($V$) frames obtained with the PC chip of the WFPC2
camera onboard the Hubble Space Telescope. }
\end{figure}


%
\begin{figure}[h]
\centerline{
\psfig{bbllx=32mm,bblly=45mm,bburx=179mm,bbury=250mm,%
figure=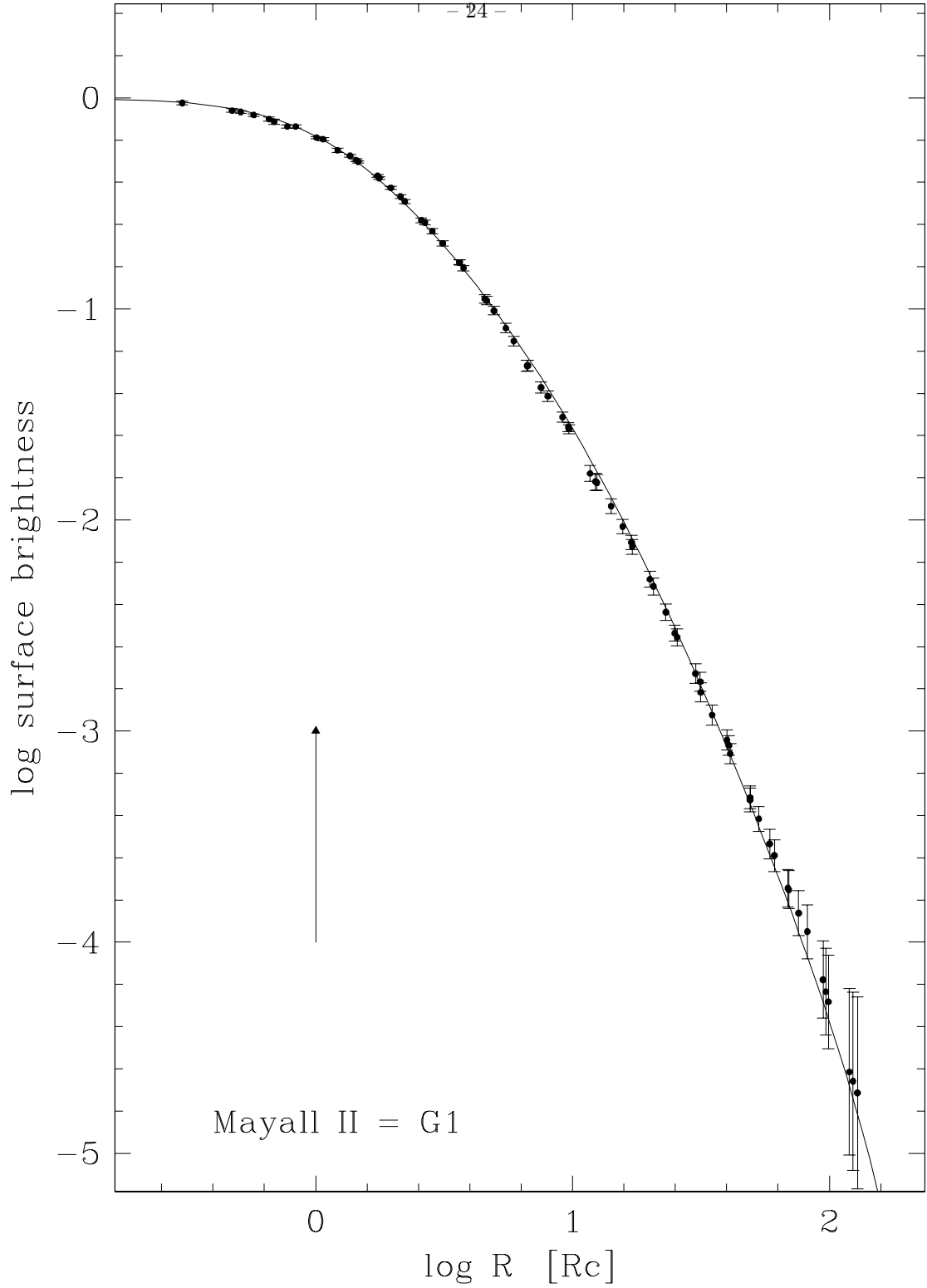,width=140mm}}
\caption[]{\label{fig.diagram}  Surface    brightness  profile  of the
globular cluster  Mayall~II $\equiv$~G1,   from HST/WFPC2 shallow  and
deep   images  in F555W   $\simeq$  $V$ filter;    the continuous line
represents a King-Michie model (first model in  Table~4) fitted to the
observed profile (Meylan \etal\ 1999). }
\end{figure}


\end{document}